\documentclass[twocolumn,showpacs,amsmath,amssymb]{revtex4}
\usepackage{graphicx}
\usepackage{float}
\usepackage{xcolor}
\usepackage{soul}

\begin{document}
\title{Shadow of the magnetically and tidally deformed black hole}
\author{R. A. Konoplya}\email{konoplya_roma@yahoo.com}
\affiliation{Research Centre for Theoretical Physics and Astrophysics, Institute of Physics in Opava, Silesian Univeristy in Opava, Bezručovo nám. 13, CZ-74601 Opava, Czech Republic}
\affiliation{Peoples Friendship University of Russia (RUDN University), 6 Miklukho-Maklaya Street, Moscow 117198, Russian Federation}
\author{Jan Schee}
\affiliation{Research Centre for Theoretical Physics and Astrophysics, Institute of Physics in Opava, Silesian Univeristy in Opava, Bezručovo nám. 13, CZ-74601 Opava, Czech Republic}
\author{Dmitriy Ovchinnikov}
\affiliation{Research Centre for Theoretical Physics and Astrophysics, Institute of Physics in Opava, Silesian Univeristy in Opava, Bezručovo nám. 13, CZ-74601 Opava, Czech Republic}

\begin{abstract}
We consider a simple model of a black hole surrounded by a toroidal structure and experiencing deformation due to magnetic field and the tidal force. For this system we construct the shadows of Preston-Poisson  black hole, apply it to supermassive black hole $M87^*$ and compare it with the shadows of corresponding Schwarzschild and Kerr black holes. We find that for large deviation parameters $B$ and $E$ the shadow of Preston-Poisson black hole is clearly distinguishable from Kerr black hole provided the angular resolution of measurements is of order $\mu\mathrm{arcsec}$.
\end{abstract}

\maketitle

\newcommand{\diff}{\mathrm{d}}

\section{Introduction}

Recent and forthcoming observations of the galactic black hole in the electromagnetic spectrum  \cite{Goddi:2017pfy,Akiyama:2019cqa} plays an important role in our understanding the black hole geometry and testing General Relativity in the strong fields regime, especially taking into account that observations in the gravitational spectrum  \cite{Abbott:2016blz,TheLIGOScientific:2016src} leaves a large room for interpretations of black hole geometry and the background gravitational theory
\cite{Konoplya:2016pmh,Wei:2018aft,Berti:2018vdi}. In this situation it is  important to complement observations of black holes in gravitational spectrum by detecting their shadows. Shadows of black holes have been recently studied in a great number of works, a big part of which could conditionally be split into two directions: shadows of black holes in alternative/modified theories of gravity \cite{Younsi:2016azx,Konoplya:2020bxa,Buoninfante:2020qud,Konoplya:2019xmn,Guo:2019lur,Konoplya:2019fpy,Zhu:2019ura,Long:2019nox,Held:2019xde,Hennigar:2018hza,Eiroa:2017uuq,Tsukamoto:2017fxq,Wang:2017hjl,Guo:2018kis,Guo:2020zmf,Allahyari:2019jqz,Stuchlik1,Stuchlik2} and shadows in presence of some distribution of matter around black holes \cite{Perlick:2017fio,Perlick:2015vta,Bisnovatyi-Kogan:2017kii,Hou:2018avu,Wang:2019tjc,Eatough:2013nva,Konoplya:2019sns,Boshkayev:2020kle,Badia:2020pnh,Pantig:2020uhp,Reji:2019brv}, to distinguish the influence of  black hole environment on observations from possible modifications of the relativity theory. One of such environmental factors of galactic black holes is a strong magnetic field in their vicinity \cite{Eatough:2013nva}.
Another essential factor representing the black hole environment is a tidal gravitational force produced by all the surrounding matter. Unlike magnetic fields, whose magnitude and decay law are  still open questions, tidal forces could be strong enough to deform (albeit minimally) the geometry of a galactic black hole as well as black holes of stellar masses having a star companion. In addition to tidal force due to visible matter near the black hole, there could be dark matter contribution as well \cite{Konoplya:2019sns}.

The straightforward way of taking into consideration  deformations of a shadow cast by the black hole immersed in such a compound environment would include analysis of a great number of matter components and would be highly dependent on the equations of state of the matter \cite{Boshkayev:2020kle}.
A great number of works were devoted to such an approach, especially when implying tidal force induced by the accreting matter around a super-massive black hole. The pioneering work was done by Luminet \cite{Luminet} who simulated a photo of the Schwarzschild black hole surrounded by a thin accretion disk. There, the primary and secondary images of the thin disk appear outside the black hole shadow. Keplerian disc and tori induced shadows of Kerr black holes were constructed in  \cite{Wu,Marck,Beckwith}. Formation of disk images and associated optical phenomena in analytically and numerically determined black hole spacetimes were also considered in \cite{Vincent}. In all those studies no deformation of the Schwarzchild or Kerr geometry was assumed. The deformed Schwarzschild geometry owing the Weyl ring was considered in \cite{Wang:2019tjc}, where the shadows were constructed for such a metric.

Here we would like to take into consideration effect of the tidal force and magnetic field onto the black hole shadow, on one hand not implying some detailed and specific origin of the tidal force, that is, for example, not being limited by accretion-disk model or extra dimensional (brane-world) tidal force, and, on the other hand, looking at such tidal force and magnetic field configurations, which are solutions of the corresponding Einstein-Maxwell equation, that is, compatible with the black hole geometry at the horizon. Therefore, a simple and robust model which allows us to understand, at least qualitatively, deformations induced by the black hole environment,  would be highly appreciated. Such a simple model, with only two deformation parameters, responsible for the tidal force and magnetic field, was suggested by Preston and Poisson \cite{Preston-Poisson,Preston:2006zd}. The basic features of particle motion in this spacetime was studied in \cite{Konoplya:2006qr} and the quasinormal modes estimated in \cite{Konoplya:2012vh}.

The Preston-Poisson P-P metric \cite{Preston-Poisson} is the perturbative solution of the Einstein-Maxwell equations with the help of the light-cone gauge formalism \cite{Preston:2006zd}. There was considered a system consisting of a black hole and a mechanical structure (a giant torus or long solenoid) at some distance around it, which was the source of the gravitational tidal force and the strong magnetic field. Thus, the background which we shall study here, the P-P space-time, has three parameters: black hole mass $M$, the asymptotic value of the strength of the magnetic field $B$ and tidal force $\mathcal{E}$.

	The black hole shadow is determined by constant $r$ photon orbits (let's denote this class of orbits by $\gamma_c$)	impact parameters. However, the Hamilton-Jacobi radial and latitudinal equations are not separable here, therefore we have to find parameters of $\gamma_c$ orbits by numerical integration of geodesics and separate the orbits passing by the black hole from those that plunge below its event horizon.
	Due to perturbative origin of P-P metric, it can be applied only near black hole. In order to b integrate photon geodesics to distant observer we need to find suitable radius $r_0$ where we tailor  P-P metric to Schwarzschild geometry. Fortunately, we are able to find photon circular orbits, $r_{ph}$, in the equatorial plane, giving us clue how to properly define radius $r_0$ so we keep the crucial information about photon orbits also in the modified spacetime.

The paper is organized as follows. We first, in section II, introduce P-P metric and discuss its applicability along with matching conditions connecting Schwarzschild spacetime with the P-P corrections. In section III we discuss construction of the black hole shadow diameter as seen from equatorial plane and compare it to Schwarzschild and Kerr cases. We also discuss ISCO orbits here. In section IV we briefly we introduce new coordinates suitable for numerical integration of geodesics. In section V, we discuss construction of P-P black hole shadow. We present plots of  $M87^*$ supermassive black hole shadows compared with Kerr and Schwarzschild black holes in section VI. The last section VII. is devoted to concluding remarks.

\section{Model of a tidally and magnetically deformed black hole}

\subsection{Preston-Poisson metric}

The P-P space-time describes the system consisting of a large mechanical structure, such as a giant solenoid or a torus which surrounds a black hole and produces an asymptotically uniform magnetic field of strength $B$. The mass of the structure is $M'$ and the radius is of order $\sim a$, while the mass of the black hole is $M$.
The electromagnetic four-vector has the following form,
\begin{equation}
A^{\mu} = \frac{1} {2} B \phi^{\mu},
\end{equation}
where $\phi^{\mu} = (0, 0, 0, 1)$. Further, it is implied that the perturbation created by the magnetic field is small, i.e.
\begin{equation}\label{cond1}
r^2 B^2 \ll 1,
\end{equation}
where $r$ is the distance from the black hole, and only the interior of the mechanical structure is under consideration $r < a$.

For our purposes the unmodified P-P metric can only be used to study the inner region of the system
which starts at the black hole horizon ($r_h = 2 M$) and ends far from the black hole, still being far from the edges of the torus,
\begin{equation}\label{cond2}
r_h \leq r \ll a.
\end{equation}
The latter condition can always be fulfilled, because the torus is supposed to be situated in the region of the weak gravitational field of the
black hole,
\begin{equation}\label{cond3}
\frac{M}{a} \ll 1.
\end{equation}
As $r^2 B^2 \ll 1$ and $r < a$, it is implied that $a^2 B^2 \ll 1$, though the relative scales of $M/a$ and $a^2 B^2$ can be arbitrary.
First of all, we are interested in the case
\begin{equation}\label{cond4}
M/a \ll a^2 B^2,
\end{equation}
i.e. in the situation when there is an asymptotic region $M \ll r < a$, where the influence of the magnetic field on the space-time geometry is negligible.
The mechanical structure of mass $M'$ produces the gravitational tidal force, parameterized by  $\mathcal{E}$, near the black hole,
\begin{equation}\label{cond5}
\mathcal{E} \sim \frac{M'}{a^3}.
\end{equation}
In the above approach the tidal force can be much larger, of the same order or much smaller than $B^2$.

Preston and Poisson used the light-cone gauge for constructing the perturbed metric, which is adapted  to incoming light cones $v =
constant$ that converge toward $r = 0$. For zero tidal force and magnetic field, $v$ takes its Schwarzschild value
$v = t + r + 2M ln(r/2M - 1)$.
In the ($v$, $r$, $\theta$, $\phi$) coordinates the P-P metric has the following form
\begin{eqnarray}\label{PPmetric1}
\diff s^2 &=& -f(r, \theta) \diff v^2 + 2p(r,\theta)\, \diff v \diff r + w(r,\theta) \diff v \diff \theta +\nonumber\\
		&& h_1(r,\theta) \diff \theta^2 + h_2(r,\theta) \diff\phi^2,
\end{eqnarray}
where is
\begin{widetext}
\begin{equation}\label{PPcomponents1}
f=1-\frac{2M}{r}+\frac{1}{9}B^2r(3r-8M)+\frac{1}{9}B^2(3r^2-14Mr+18M^2)(3\cos^2\theta-1)    -\mathcal{E}(r-2M)^2(3\cos^2\theta-1)+O[B^4,\mathcal{E}^2],
\end{equation}
\begin{equation}\label{PPcomponents2}
p=1+O[B^4,\mathcal{E}^2]\text{,}
\end{equation}
\begin{equation}\label{PPcomponents3}
w=\frac{2}{3}B^2r^2(r-3M)\sin\theta\cos\theta-2\mathcal{E} r^2(r-2M)\sin\theta\cos\theta+O[B^4,\mathcal{E}^2],
\end{equation}
\begin{equation}\label{PPcomponents4}
h_1=r^2-\frac{2}{9}B^2r^4+\frac{1}{9}B^2r^4(3\cos^2\theta-1)+B^2M^2r^2\sin^2\theta+\mathcal{E} r^2(r^2-2M^2)\sin^2\theta+O[B^4,\mathcal{E}^2],
\end{equation}
\begin{equation}\label{PPcomponents5}
h_2=r^2\sin^2\theta-\frac{2}{9}B^2r^4\sin^2\theta+\frac{1}{9}B^2r^4\sin^2\theta(3\cos^2\theta-1)-B^2M^2r^2\sin^4\theta -\mathcal{E} r^2(r^2-2M^2)\sin^4\theta+O[B^4,\mathcal{E}^2].
\end{equation}
\end{widetext}
The above metric was obtained in \cite{Preston-Poisson} by the perturbation of the Enstein-Maxwell equation in orders of $(\mathcal{E}, B^2)$.
The parameter $\mathcal{E}$  is the Weyl curvature, that is, the tidal
gravitational field, of the asymptotic space-time measured
by an observer co-moving with the black hole in the
region $M \ll r\ll 1/B$.
The perturbed event horizon is given by
\begin{equation}
r_{h} = 2 M (1+ \frac{2}{3} M^2 B^2 \sin^2 \theta).
\end{equation}
It is essential
that the event horizon is affected by $B$ and not by $\mathcal{E}$.

Although in the above space-time only the dominant order of the magnetic field is considered, the same relation as for the Ernst black holes takes place
\begin{equation}
B \sim 10^{-21} \frac{M}{M_{\odot}} B_{0},
\end{equation}
where $M$ and $M_{\odot}$ are the mass of a black hole and of the sun respectively and $B_{0}$ is the external magnetic field in units of gauss.
From the above relation one can see that the magnetic field deforming the space-time geometry significantly would be as strong as $B = 10^{12} G$ for galactic black holes with  mass $M \sim 10^{9} M_{\odot}$.

\subsection{The matching conditions}

As the original perturbative metric is valid only as soon as $r ^2 B^2 \ll 1$ and $r^2 \mathcal{E} \ll 1$, we cannot apply it large values of $B$ and $\mathcal{E}$ and have an observer situated sufficiently far from the black hole. Therefore, it is reasonable to match this perturbative solution with the one which would provide sufficiently quick decay of the tidal force and magnetic field at infinity, leading to the post-Newtonian behavior at large distance, which is compatible with the current observations. For this we start with the transformations
\begin{equation}
v = t + F(r), \quad -F'(r) f(r, \theta) + 1 =0
\end{equation}
reduce the metric (\ref{PPmetric1}) to the following form
\begin{eqnarray}\label{reduced-metric}
\diff s^2 &=& -f(r, \theta) \diff t^2 + f(r, \theta)^{-1} \diff r^2 + h_1(r,\theta) \diff \theta^2+\nonumber\\
	&&   h_2(r, \theta) \diff\phi^2 + w(r, \theta) (\diff t + \frac{\diff r}{f(r,\theta)}) \diff \theta.
\end{eqnarray}
Then, we can observe that
\begin{equation}
g_{\mu \nu} = g_{\mu \nu}^{Schw} + B^2 g_{\mu \nu}^{1} (\theta, r) + \mathcal{E} g_{\mu \nu}^{2} (\theta, r),
\end{equation}
where $g_{\mu \nu}^{Schw}$ represents components of the Schwarzschild metric.
The highest power of $r$ in the asymptotic behavior of the functions $g_{\mu \nu}^{1} $ and $g_{\mu \nu}^{2} $ is four:
\begin{equation}
g_{\theta \theta}^{1} \sim g_{\theta \theta}^{2} \sim r^4, \quad r \rightarrow \infty.
\end{equation}
Then, we require that at some value of the radial coordinate $r_{0}$, such that still we have $r_{0} ^2 B^2 \ll 1$ and $r_{0}^2 \mathcal{E} \ll 1$ the P-P metric is smoothly matched with some metric $\tilde{g}_{\mu \nu}$ interpolating between the Schwazrschild-like asymptotically flat spacetime and the P-P metric. Therefore, we imply that
\begin{equation}\label{fit1}
g_{\mu \nu}\biggr|_{r=r_{0}}= \tilde{g}_{\mu \nu}\biggr|_{r=r_{0}}
\end{equation}
and
\begin{equation}\label{fit2}
\frac{\partial g_{\mu \nu}}{\partial x^{\rho}}\biggr|_{r=r_{0}}= \frac{\partial\tilde{g}_{\mu \nu}}{\partial x^{\rho}}\biggr|_{r=r_{0}}.
\end{equation}
In order to provide the correct post-Newtonian behavior, it is sufficient to require that interpolating metric have the similar form to the P-P one, but the tidal force and magnetic field become the following decaying functions of $r$
\begin{equation}
B^2 \rightarrow \frac{\beta_{1}}{r^5} +\frac{\beta_{2}}{r^{6}},
\end{equation}
\begin{equation}
\mathcal{E} \rightarrow \frac{\epsilon_{1}}{r^5} +\frac{\epsilon_{2}}{r^{6}}.
\end{equation}
From the matching conditions (\ref{fit1}) and (\ref{fit2}), we find that
\begin{equation}
\beta_{1} = \frac{6}{11} r_{0}^5 B^2, \quad \beta_{2} = - \frac{5}{11} r_{0}^6 B^2,
\end{equation}
\begin{equation}
\epsilon_{1} = \frac{6}{11} r_{0}^5 \mathcal{E}, \quad \epsilon_{2} = - \frac{5}{11} r_{0}^6 \mathcal{E}.
\end{equation}
With these values at hand we are ready to consider the metric which is valid everywhere in the space outside the event horizon until infinity.

\section{Equations of motion and the effective potential}

We first discuss test particle motion in equatorial plane $\theta=\pi/2$ where  P-P spacetime simplifies to
\begin{equation}
\diff s^2=-f(r,\pi/2)\diff t^2+\frac{\diff r^2}{f(r,\pi/2)}+h_2(r,\pi/2)\diff\phi^2.
\end{equation}
We discuss the angular diameter of the black hole as seen from equatorial plane and construct the ISCO orbits.

\subsection{ Black-hole shadow diameter in equatorial plane}
Let us compare the shadows of Schwarzchild, Kerr ($a=0.9982$), and PP black hole shadows diameter in the equatorial plan. Black hole shadow diameter is given by the value of impact parameter $l_{ph}$ of photon circular orbit. Considering a black hole at distance $d_0$, then its angular diameter will read
\begin{equation}
	\alpha = 2\,\arcsin\frac{l_{ph}}{d_0}.\label{diameter}
\end{equation}
Let the mass of the black hole be $M=M_{M87*}\approx 6\times 10^9 M_\odot$ and its distance from observer is $d_o=16.2$Mpc. We calculate its angular diameter for three models:
\begin{itemize}
	\item \emph{Schwarzchild metric} - photon orbit is located at $r_{ph}=3 G M/c^2$ and corresponding impact parameter reads $l_{ph}=3\sqrt{3} G M/c^2$. Inserting into (\ref{diameter}) we obtain Schwarzchild black hole angular diameter
	\begin{equation}
		\alpha_{\textrm{Schw}}=82.6\,\mu\textrm{arcsec}
	\end{equation}.
	\item \emph{Kerr metric} - the radial equation for photon geodesics in Kerr in equatorial plane reads
	\begin{eqnarray}
		(k^u)^2&=&U(u;a,l)\nonumber\\
			   &=&1+(a^2-l^2)u^2+2(a-l)^2 u^3
	\end{eqnarray}
	implying that photon circular orbits are solution of equations
	\begin{equation}
		U(u;l,a)=0\quad\textrm{and}\quad \frac{\diff U}{\diff u}=0
	\end{equation}
	leading to equation for impact parameter $l$ in form
	\begin{eqnarray}
		x^3-27x+54a=0
	\end{eqnarray}
	where we have introduced new variable $x\equiv a+l$. For spin $a=0.9982$ corresponding impact parameter has value $l_{ph}=2.105132 G M/c^2$. The angular diameter of the Kerr black hole has value
	\begin{equation}
		\alpha_{\textrm{Kerr}}=33.48\,\mu\textrm{arcsec}.
	\end{equation}
	\item \emph{P-P metric} - here, photon circular orbits are solutions of equations
	\begin{equation}
		f'(r)h_2(r)-f(r)h_2'(r)=0\quad\Rightarrow\quad r_{ph}
	\end{equation}
	and
	\begin{equation}
		l_{ph}=\frac{h_2(r_{ph})}{f(r_{ph})}.
	\end{equation}
	Again, considering black hole mass $M=M_{M87*}$, distance $d_0$ and magnetic field parameters to be  $B=0.1$ and $E=0.01$ then the impact parameter of photon circular orbit and corresponding diameter of its shadow will reach value $l_{ph}=4.723675$ and
	\begin{equation}
		\alpha_{\textrm{PP}}=75.14\,\mu\textrm{arcsec}.
	\end{equation}
\end{itemize}

\subsection{Innermost Stable Circular Orbits}
\begin{figure}[H]
	\begin{center}
		\includegraphics[scale=0.6]{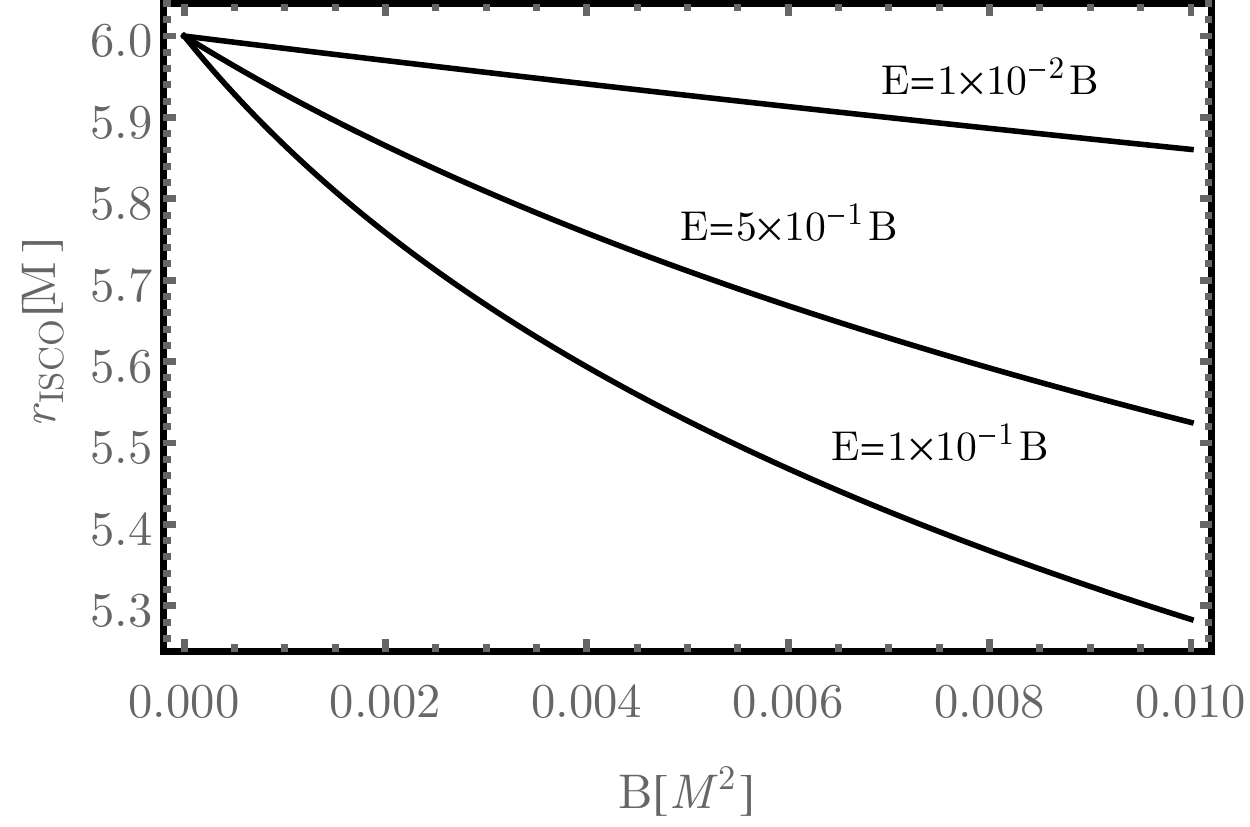}
		\caption{\label{pp_isco} Plot of $r_{ISCO}$ as function of magnetic field parameter $B$ for three representative relations between $B$ and $E$.}
	\end{center}
\end{figure}

We consider an observer moving in equatorial plane ($\theta=\pi/2$) along a circular orbit (we will mainly consider ISCO). The spacetime interval  (\ref{reduced-metric}) simplifies to read
\begin{equation}
\diff s^2=-f(r,\pi/2)\diff t^2+\frac{\diff r^2}{f(r,\pi/2)}+h_2(r,\pi/2)\diff\phi^2.
\end{equation}
Clearly, there are two constants of motion $p_t=-E$ and $p_\phi=L$ associated with the symmetries of the spacetime, then equations of motion of test particle can be written in the form
\begin{eqnarray}
	\left(u^r\right)^2&=&E^2 - V_{eff}(r),\\
	u^t&=&\frac{E}{f(r,\pi/2)},\\
	u^\phi&=&\frac{L}{h_2(r,\pi/2)}
\end{eqnarray}
where we have introduced the effective potential
\begin{equation}
	V_{eff}=f(r,\pi/2)\left(1+\frac{L^2}{h_2(r,\pi/2)}\right).
\end{equation}
Moving along circular geodesics, the conditions
\begin{equation}\label{circ_orb}
	E^2=V_{eff}\quad\textrm{and}\quad V'_{eff}=0.
\end{equation}
Considering ISCO (sometimes called marginally stable orbit) another condition must be satisfied
\begin{equation}\label{isco}
	V''_{eff}=0.
\end{equation}
must be satisfied. Conditions (\ref{circ_orb}) and (\ref{isco}) lead to set of two equations for two unknown parameters $L^2$ and $r=r_{ISCO}$, reading
\begin{eqnarray}
	L^2&=&\frac{h_2}{f'}\left(f\frac{h'_2}{h_2}-f'\right)^{-1},\\
	L^2&=&f''\left[\frac{f}{h_2^2}\left(h''_2 -\frac{2h''_2\phantom{}^2}{h_2}\right)+\frac{2f'h'_2}{h_2^2}-\frac{f''}{h_2}\right]^{-1}.
\end{eqnarray}
The loci of ISCO as function of magnetic field $B$ is presented in Fig. \ref{pp_isco}.

\section{Raytracing}
In order to connect observer's detector with distant source on the sky we use raytracing technique.
For numerical purpose it is convenient to transform spacetime (\ref{reduced-metric}), using new coordinates $u=1/r$ and $m=\cos\theta$, to read
\begin{eqnarray}\label{num_metric}
		\diff s^2&=&-\tilde{f}(u,m)\diff t^2 + \frac{\diff u^2}{u^4\tilde{f}(u,m)} + \frac{\tilde{h}_1(u,m)\diff m^2}{1-m^2}\nonumber\\
		&&+\tilde{h}_2(u,m)\diff\phi^2\nonumber\\
	&&-\frac{\tilde{w}(u,m)}{\sqrt{1-m^2}}\left(\diff t -\frac{\diff u}{u^2\tilde{f}(u,m)}\right)\diff m
\end{eqnarray}
where is $\tilde{f}(u,m)=f(1/u,\arccos(m))$, $\tilde{h}_1(u,m)=h_1(1/u,\arccos(m))$, $\tilde{h}_2(u,m)=h_2(1/u,\arccos(m))$, and $\tilde{w}(u,m)=w(1/u,\arccos(m))$.
Equation's of motion are not separable, therefore we construct geodesic equations
\begin{equation}
	\frac{\diff k^\mu}{\diff \lambda}+\Gamma^\mu_{\phantom{\mu}\alpha\beta}k^\alpha k^\beta=0\label{geodesics}
\end{equation}
out of the metric (\ref{num_metric}), where is $k^\mu$ the components of wave propagation vector $\vec{k}$. In order to numerically solve the geodesic equation we have to specify initial position vector $x_0^\mu$ and initial components of propagation vector $k_0^\mu$. We discuss the choice of $x_0^\mu$ and $k_0^\mu$ in the following section.

\section{Construction of the Shadow}
\begin{figure}[H]
	\begin{center}
		\includegraphics[scale=0.5]{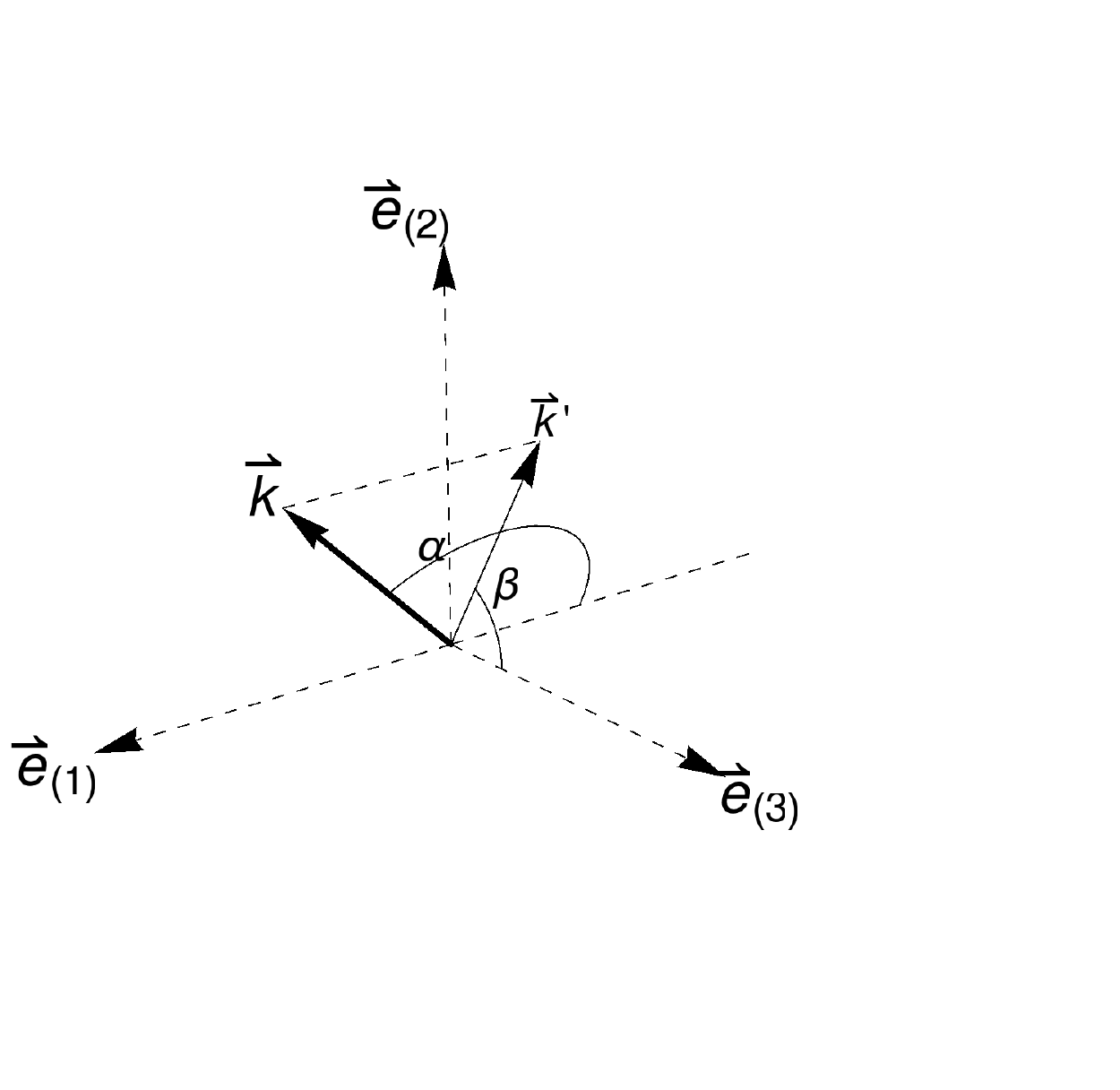}
		\caption{\label{vectors1}Illustration of tetrade vectors and directional angles $\alpha$ and $\beta$ of propagation vector $\vec{k}$ definition. Usefull for construction of black hole shadow allone.}
	\end{center}
\end{figure}

\begin{figure}[H]
	\begin{center}
		\includegraphics[scale=0.5]{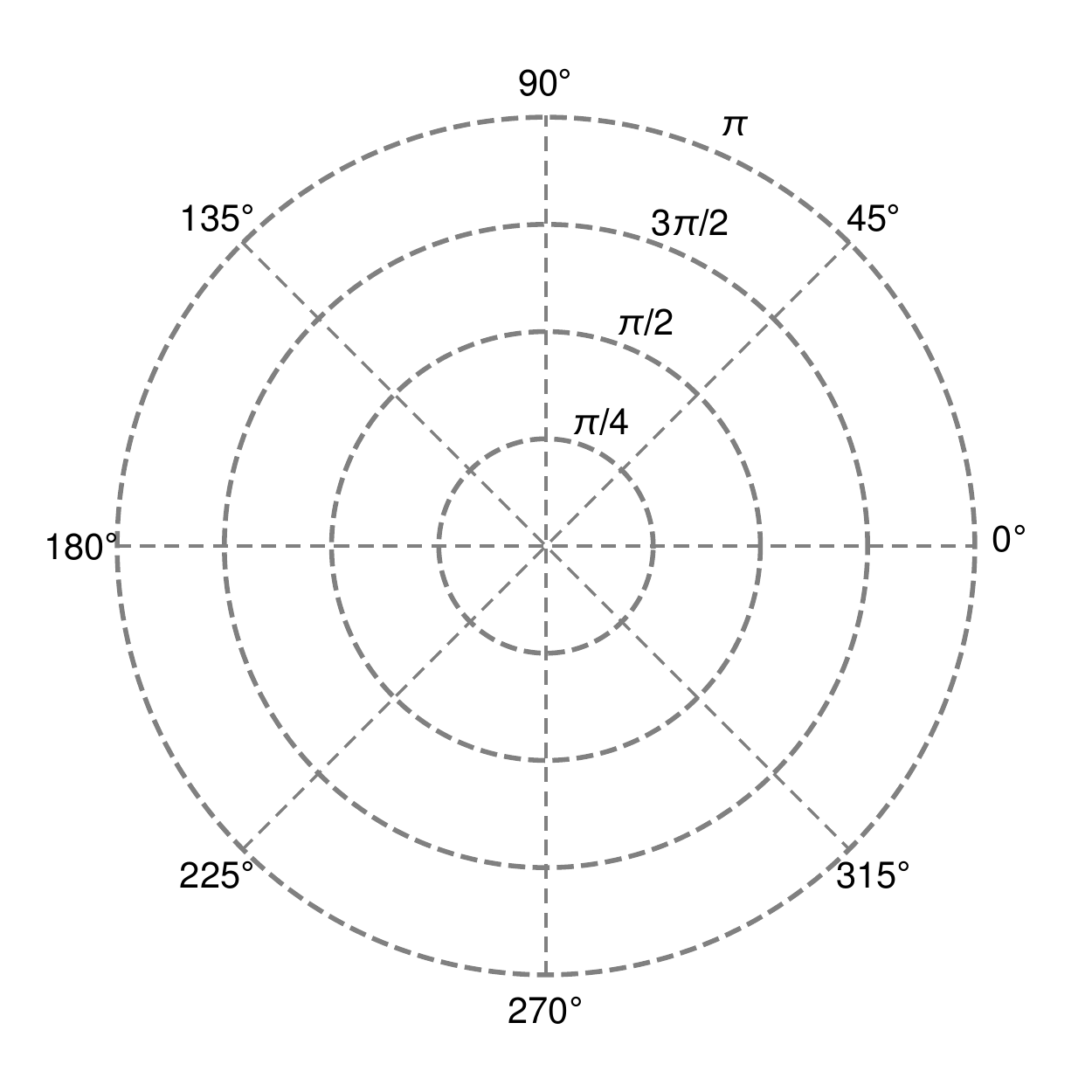}
		\caption{\label{polar}Polar plot grid. Radial lines represent angle $\alpha$ and circles represent angle $\beta$.}
	\end{center}
\end{figure}

\begin{figure}[H]
	\begin{center}
		\includegraphics[scale=0.5]{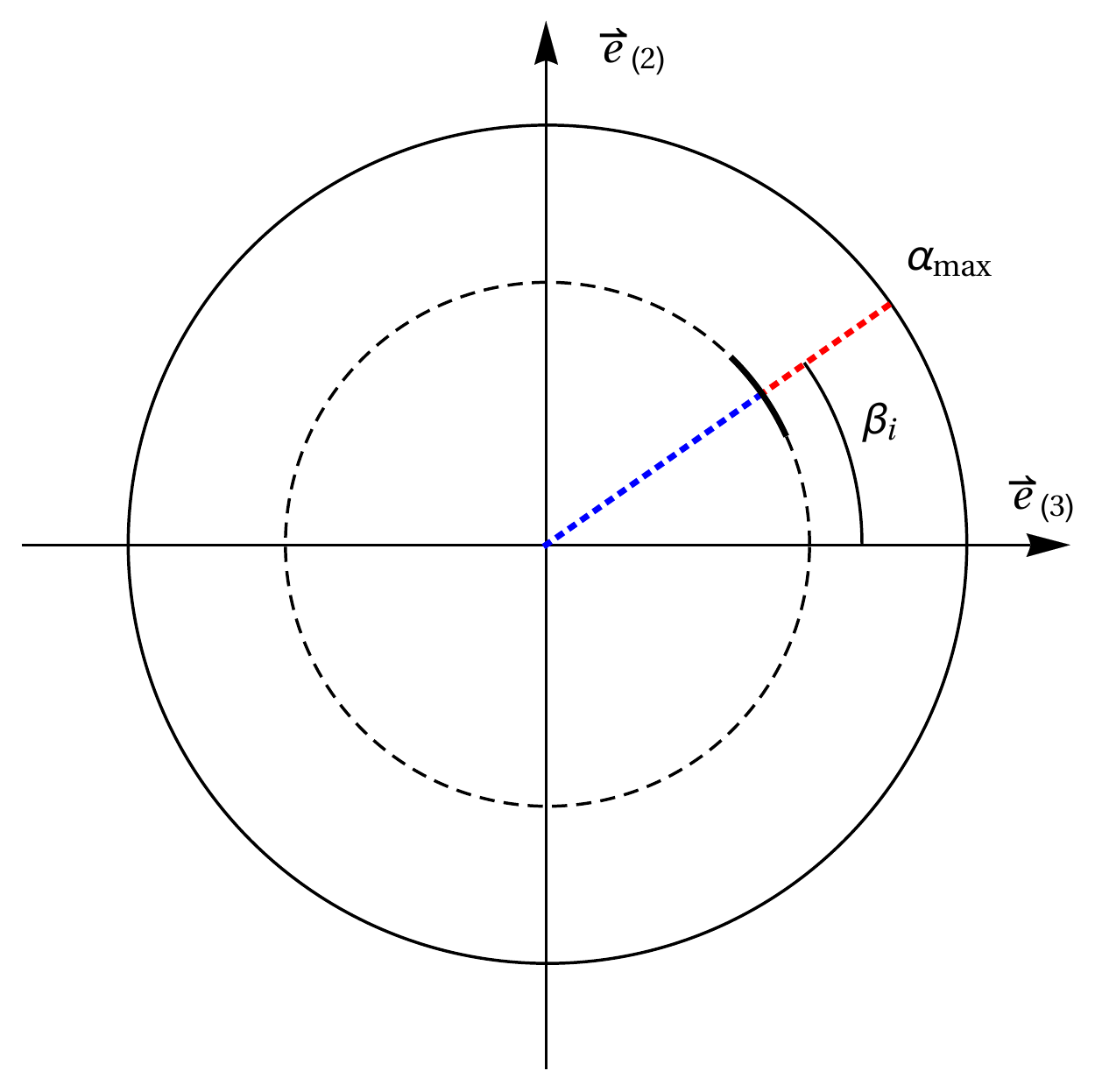}
		\caption{\label{algorithm}}
	\end{center}
\end{figure}

The shadow of the black hole is given by total, photon orbit impact parameter defined by formula
\begin{equation}
	b_{ph}\equiv\sqrt{l_{ph}^2+q_{ph}}.
\end{equation}
The corresponding diameter of the black hole shadow is simply calculated from the following equation
\begin{equation}
	\alpha_{M87}=2\,\arcsin\frac{b_{ph}G\,M}{c^2\,d_0}.\label{alphaM87}
\end{equation}
In order to obtain $b_{ph}$, consider a tetrade of static Minkowski observer in the form
\begin{eqnarray}
	\vec{e}_{(0)}&=&\vec{e}_t,\\
	\vec{e}_{(1)}&=&\vec{e}_r,\\
	\vec{e}_{(2)}&=&\frac{1}{r_o}\vec{e}_\theta=- u_o\sqrt{1-m_o^2}\vec{e}_\theta\\
	\vec{e}_{(3)}&=&\frac{1}{r_o\sin\theta_o}\vec{e}_{\phi}=\frac{u_o}{\sqrt{1-m_o^2}}\vec{e_{\phi}}.
\end{eqnarray}
In order to construct the black hole shadow alone, as seen by distant observer, we redefine the directional angles $\alpha$ and $\beta$ to read (see also Fig \ref{vectors1})
\begin{eqnarray}
k^{(0)}&=&-k_{(0)}=1,\\
k^{(1)}&=&k_{(1)}=-\cos\alpha,\\
k^{(2)}&=&k_{(2)}=\sin\alpha\,\sin\beta,\\
k^{(3)}&=&k_{(3)}=\sin\alpha\,\cos\beta.	
\end{eqnarray}
The resulting black hole shadow is represented by the polar plot as shown in Fig. \ref{polar}.
The relationship between photon propagation 4-vector components relative to coordinate base $k^\mu$ and tetrade $k^{(a)}$ read
\begin{eqnarray}
	k^t &=& k^{(0)}=1,\label{kt0}\\
	k^u &=& - u_o^2\,k^r=-uo^2\,k^{(1)}=-u_o^2\,\cos\alpha,\label{ku0}\\
	k^m &=& - \sqrt{1-m_o^2}k^\theta=-u_o\sqrt{1-m_o^2}k^{(2)}\nonumber\\
		&=&-u_o\sqrt{1-m_o^2}\sin\alpha\,\sin\beta,\label{km0}\\
	k^\phi &=&\frac{k^\phi}{r_o\,\sin\theta_o}=\frac{u_o}{\sqrt{1-m_o^2}}k^{(3)}\nonumber\\
		   &=&\frac{u_o}{\sqrt{1-m_o^2}}\sin\alpha\,\cos\beta\label{kp0}.
\end{eqnarray}
Now, we are ready to express impact parameters $l$ and $q$ in terms of angles $\alpha$ and $\beta$. We have
\begin{eqnarray}
	l&\equiv&-\frac{k_\phi}{k_t}=\frac{\sqrt{1-m_o^2}}{u_o}\sin\alpha\,\cos\beta,\label{lph}\\
	q&=&\frac{\sin^2\alpha}{u_o^2}\left[1-\left(1-m_o^2\right)\cos^2\beta\right].\label{qph}
\end{eqnarray}
Figure \ref{algorithm} illustrates the algorithm of finding boundary of the black hole shadow.
We follow the steps:
\begin{itemize}
	\item Divide the interval $I_\beta=[0,2\pi]$ into $N$ grid points. Each gridpoint $\beta_i$ is determined by the rule
	\begin{equation}
		\beta_i=i\,\frac{2\pi}{N}.
	\end{equation}
	\item Divide interval $I_\alpha=[0,\alpha_{max}]$ into $M$ grid points. Each gridpoint $\alpha_j$ is determined by the rule
	\begin{equation}
			\alpha_j=j\,\frac{\alpha_{max}}{M}.
	\end{equation}
	\item For each $\alpha_j$ determine initial conditions (\ref{kt0}) - (\ref{kp0}).
	\item Integrate system of differential equations until you reach either horizon $r_h$ or turning point, i.e. the moment where $u_{i+1}<u_i$.
	\item Let us identify the geodesics ending up at horizon with $-1$ (blue color in Fig. \ref{algorithm}) and  those possessing a turning point with $1$ (red color in Fig. \ref{algorithm}). Let us denote by index $j$ the first grid point corresponding to red points. The value of $\alpha_s$ for the shadow boundary falls into the interval $[\alpha_{j-1},\alpha_{j}]$.
	\item From equations (\ref{lph}) and (\ref{qph}) determine corresponding $l_{ph}$ and $q_{ph}$ using values $\beta_i$ and $\alpha_s$.
	\item For calculated $l_{ph}$ and $q_{ph}$ determine $\alpha_{M87}$ from formula (\ref{alphaM87}).
\end{itemize}

\section{Results}
Following the presented analysis we have constructed shadows of the supermassive black hole $M87^*$ with mass $M_{M87}\approx 6.5\times 10^{9}\mathrm{M}_{\odot}$ and being at the distance $d_0=16.2\mathrm{Mpc}$.
\begin{figure}[H]
	\begin{center}
		\begin{tabular}{c}
		\includegraphics[scale=0.6]{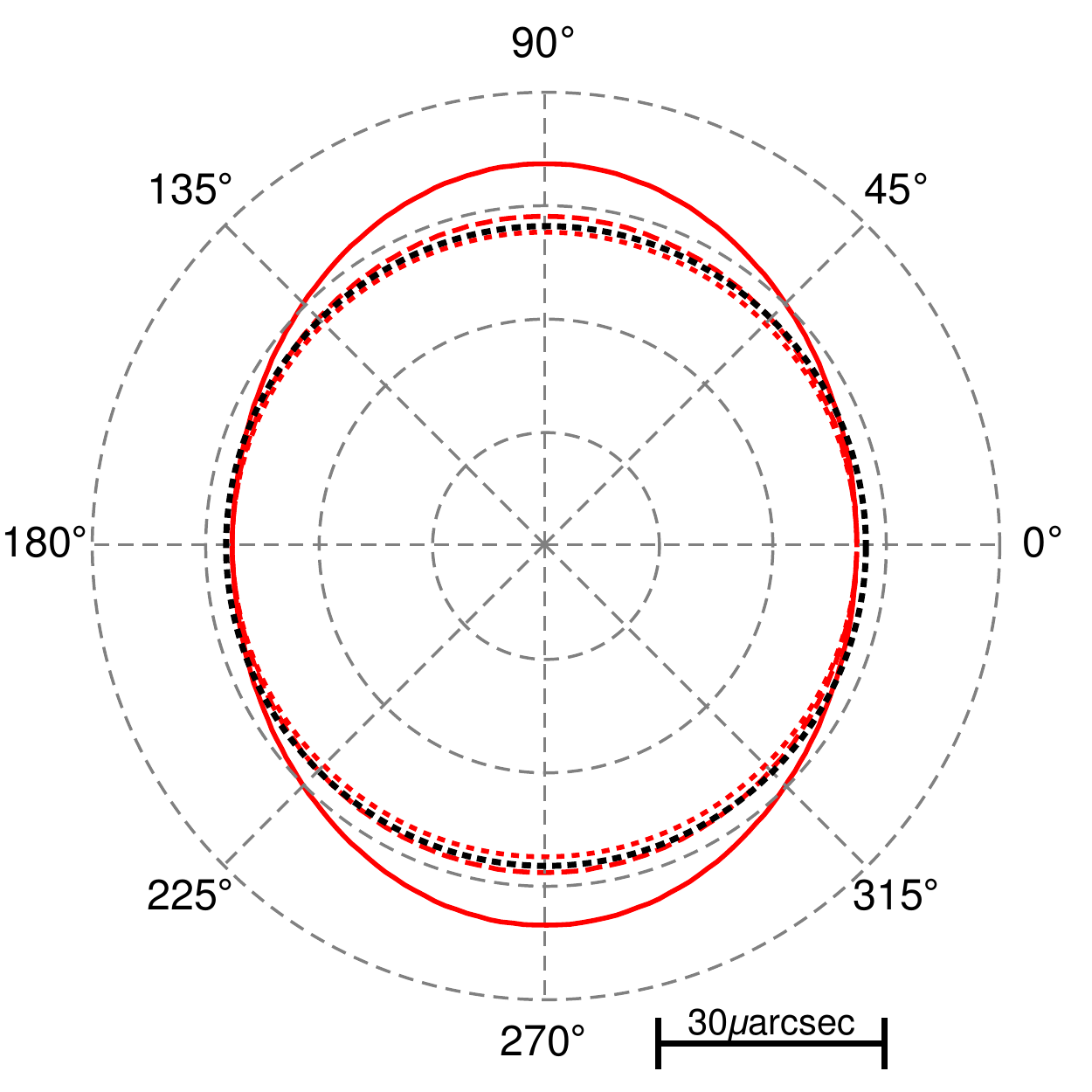}
		\end{tabular}
		\caption{The P-P black hole shadow border of supermassive black hole residing at the center of M87 galaxy constructed for three representative values of $(B,\mathcal{E})=(10^{-3},10^{-4})$ (red,solid), $(5\times 10^{-4},2.5\times 10^{-5})$ (red,dashed) and $(10^{-5},10^{-6})$ (red, dotted). For reference we draw also Schwarzchild black hole shadow border (black, dotted). }\label{fig_shadow1}
	\end{center}	
\end{figure}

\begin{figure}[H]
	\begin{center}
		\begin{tabular}{c}
			\includegraphics[scale=0.6]{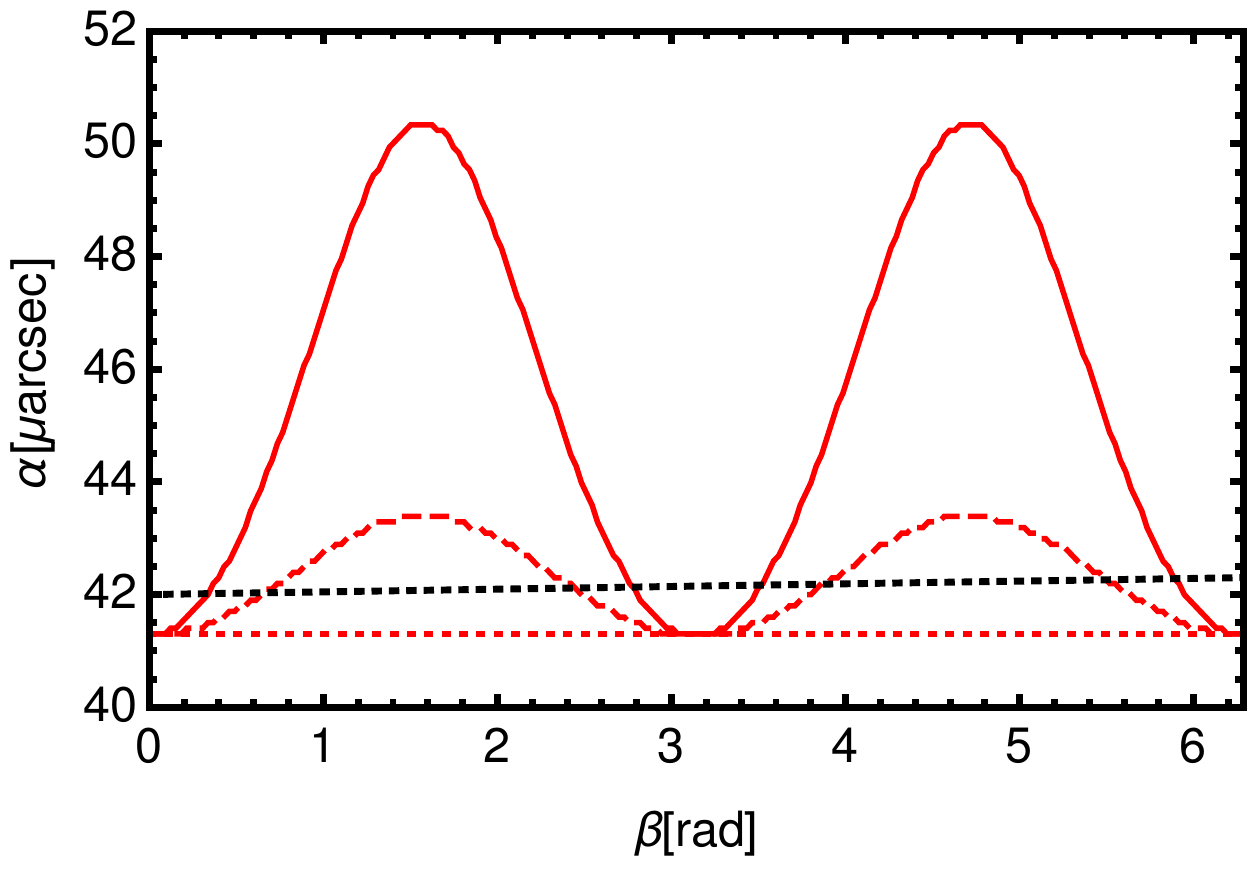}
		\end{tabular}
		\caption{The PP-black hole shadow border radius as function of angle $\beta$ for supermassive black hole residing at the center of M87 galaxy constructed for three representative values of$(B,\mathcal{E})=(10^{-3},10^{-4})$ (red,solid), $(5\times 10^{-4},2.5\times 10^{-5})$ (red,dashed) and $(10^{-5},10^{-6})$ (red, dotted). For reference we draw also Schwarzschild black hole shadow border (black, dotted). }\label{fig_cmp_1}
	\end{center}	
\end{figure}

\begin{figure}[H]
	\begin{center}
		\includegraphics[scale=0.6]{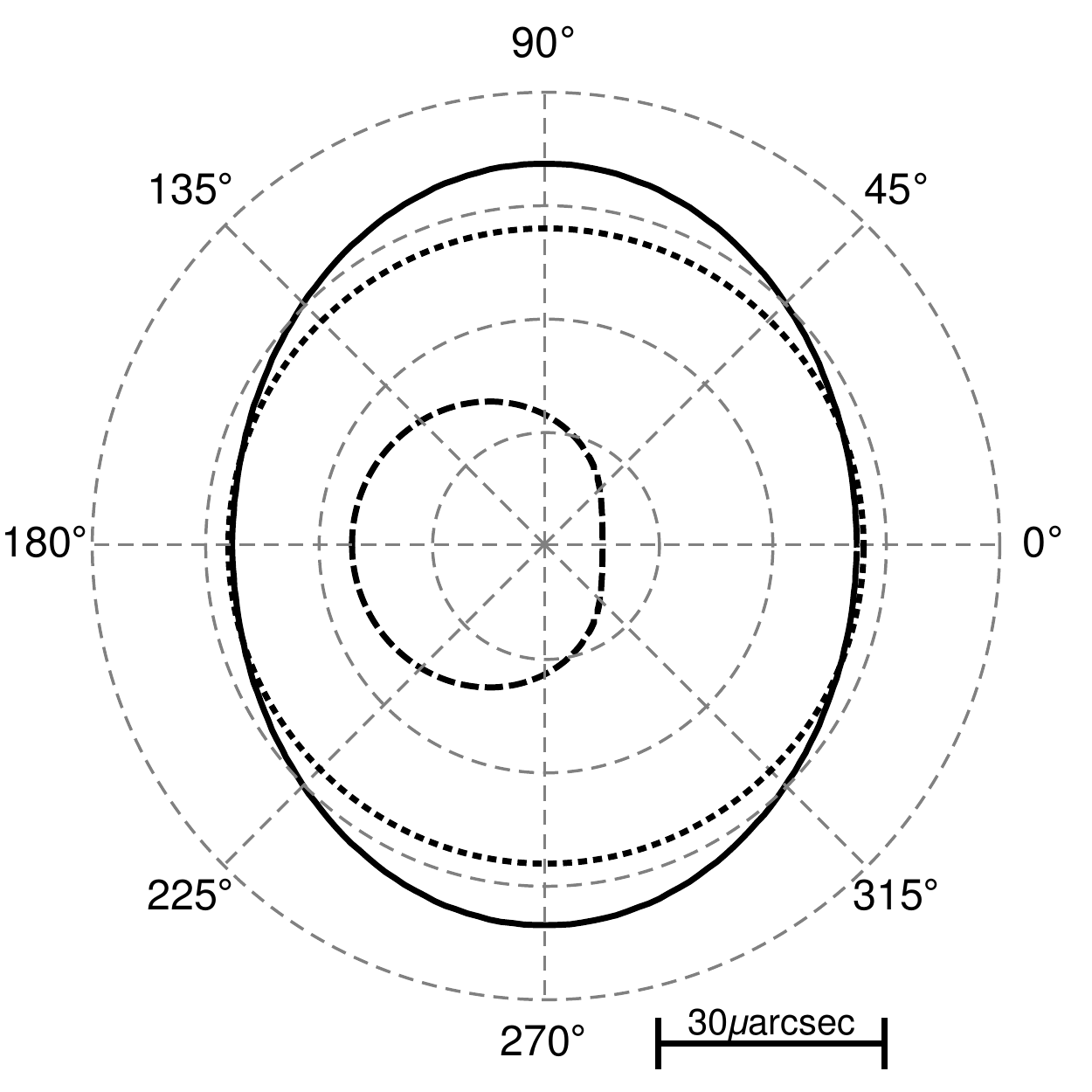}
		\caption{The black hole shadow border of supermassive black hole residing at the center of M87 galaxy constructed for P-P black hole with spacetime parameters set to $(B,\mathcal{E})=(10^{-3},10^{-4})$ (solid), Kerr black hole with spin $a=0.98$ (dashed), and Schwarzschild black hole (dotted). }\label{fig_shadow2}
	\end{center}
\end{figure}

\begin{figure}[H]
	\begin{center}
		\includegraphics[scale=0.6]{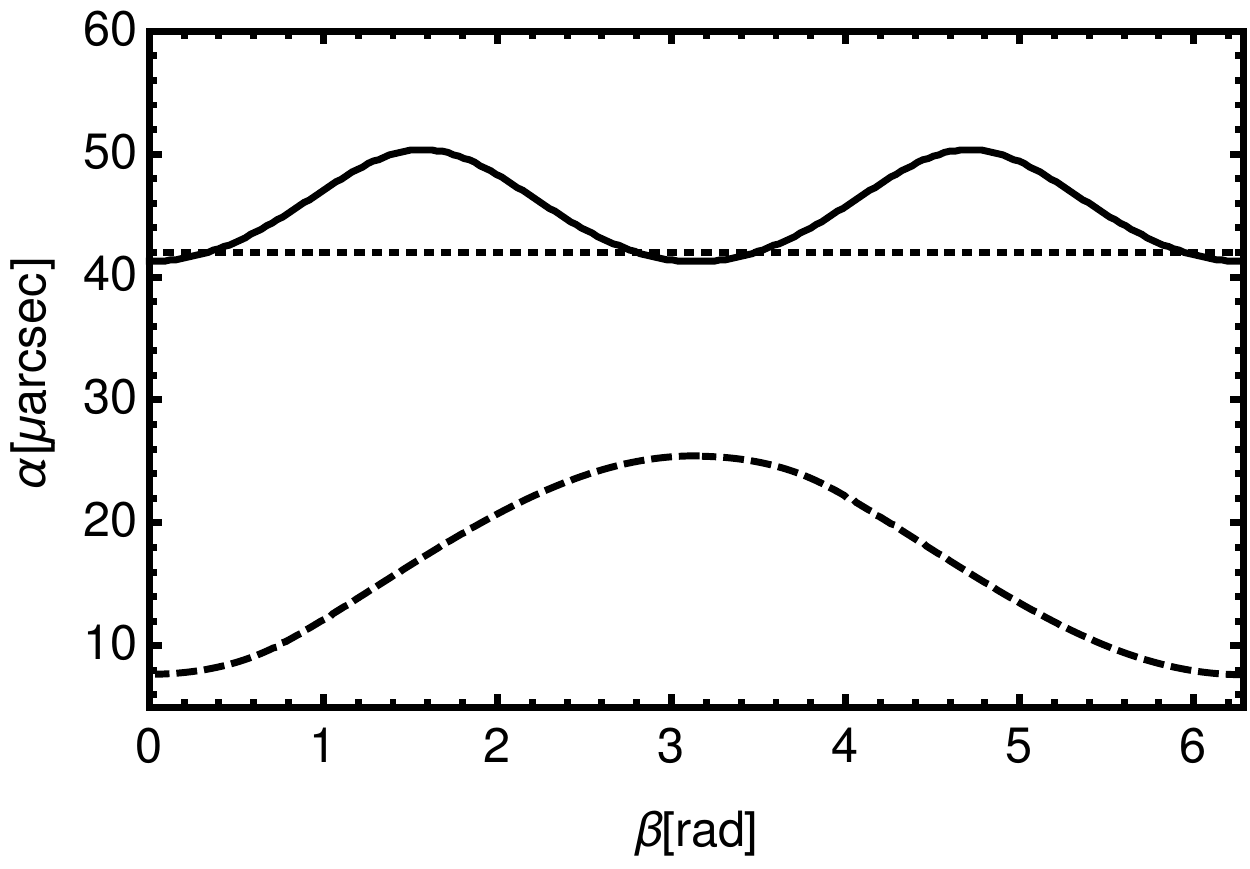}
		\caption{The black hole shadow border radius $\alpha$ as a function of angle $\beta$ for supermassive black hole residing at the center of M87 galaxy constructed for P-P black hole with spacetime parameters set to $(B,\mathcal{E})=(10^{-3},10^{-4})$ (solid), Kerr black hole with spin $a=0.98$ (dashed), and Schwarzschild black hole (dotted).}\label{fig_cmp_2}
	\end{center}
\end{figure}

\section{Conclusion}

We have studied the shadow cast by the deformed Schwarzschild black hole, supposing that deformations were induced by the tidal force and magnetic field and represented by the P-P solution \cite{Preston-Poisson}. We have shown that even relatively small values of the tidal force or magnetic field produce noticeable deviation of the shadow from spherical shape, making the shadow prolate.

In our simulations observer is located in the equatorial plane of  the black hole. We have first discussed diameter of the P-P black hole shadow considering photons moving in equatorial plane only. There we can have constructed effective potential of null geodesics and have found photon circular orbits. This result gives us clue how to choose radius $r_0$ to tailor Schwarzchild spacetime to P-P spacetime. Denoting $r_{ph}$ the radius of photon circular orbit, we have chosen $r_0=1.5r_{ph}$. The resulting diameter of the P-P black hole ($B=0.1$, $E=0.01$) shadow is  $75.14\mu\mathrm{arcsec}$ and is smaller then the corresponding shadow of Schwarzschild black hole ($41.3\mu\mathrm{arcsec}$) and it is larger than the shadow of corresponding Kerr black hole ($a=0.9982$) being $33.48\mu\mathrm{arcsec}$.

Numerical integration of geodesics equations for  initial conditions (\ref{ku0})-(\ref{kp0}) we have constructed full shadows of P-P black holes for three representative values of spacetime parameters $B$ and $E$ presented in Figs \ref{fig_shadow1}. and \ref{fig_cmp_1}. There, one can see that the larger are the values of spacetime parameters $B$ and $E$, the smaller is the horizontal (line $0^\circ$ - $180^\circ$) diameter of P-P black hole shadow. On the other hand the reverse behavior holds for vertical (line $90^\circ$ - $270^\circ$) diameter of the shadow.

We have also compared the P-P black hole shadow to Kerr black hole shadow having spin $a=0.9982$, see Figs. \ref{fig_shadow2} and \ref{fig_cmp_2}. For large spin $a$ and large values of $B$ and $E$ the P-P and Kerr black holes shadows deviate from circular Schwarzchild shape significantly and one can clearly tell apart P-P and Kerr black holes shadows in this case. On the other hand both, P-P and Kerr black hole shadows become indistinguishable from Schwarzschild shadow for $a$, $B$ and $E$ parameters approaching zero.

Here we have shown that the deformation of the black hole shadow due to the tidal force and/or magnetic field is clearly distinguishable from the effect of rotation when comparing with the Kerr shadow seen by a distant observer. Nevertheless, having in mind that slowly rotating black hole could also be observed in the future and, the black hole geometry could potentially be described by modified gravitational theory for which, in principle, any kind of deformations are allowed, it is important to understand, first, the contribution of the black hole environment to spacetime geometry (and subsequently to formation of  the black hole shadow) in order to distinguish it from that stipulated by a possible alternative theory of gravity. The presented work was the first step in this direction where the P-P perturbative metric was used as a black hole model.
Then, it would be interesting to generalize our work and construct a shadow of the deformed Kerr black hole found in \cite{Poisson:2014gka,Chatziioannou:2016kem}, which however would be much more involved and, unlike the simple and robust case considered here, depends on a great number of tidal parameters.

\acknowledgments{This work was supported by the 9-03950S GAČR grant.}

\appendix
\section{Preston-Poisson Shadow}
Here we illustrate possible shapes of the P-P shadow as seen by static frame (SF) and circular orbit frame (COF) observers in the close vicinity of the P-P black hole.
\subsection{Static Frame Tetrad}
The static observer 4-velocity reads
\begin{equation}
\vec{u}=\frac{1}{\sqrt{f}}\vec{e}_t.
\end{equation}
We naturally set the temporal tetrad vector $\vec{e}_{(0)}$ to observer 4-velocity $\vec{u}$, i.e.
\begin{equation}
\vec{e}_{(0)}=\vec{u}=\frac{1}{\sqrt{f}}\vec{e}_t.\label{tr0a}
\end{equation}
The remaining three tetrad vectors then read
\begin{eqnarray}
\vec{e}_{(1)}&=&\sqrt{f}\vec{e}_r,\label{tr1a}\\
\vec{e}_{(2)}&=&A\,\vec{e}_t + B\,\vec{e}_r+C\,\vec{e}_\theta,\label{tr2a}\\
\vec{e}_{(3)}&=&\frac{1}{\sqrt{h_2}}\vec{e}_{\phi}\label{tr3a}
\end{eqnarray}
where is
\begin{eqnarray}
A &=& \frac{w}{\sqrt{H\,f}},\\
B &=& -\frac{w\sqrt{f}}{\sqrt{H}},\\
C &=& \frac{2\sqrt{f}}{\sqrt{H}}
\end{eqnarray}
with
\begin{equation}
H=4f\, h_1.
\end{equation}
\subsection{Circular Orbit Frame Tetrad}
One can easily show that Lorentz transformation $\Lambda_{(a')}^{\phantom{a}(b)}$ of a tetrad $\vec{e}_{(a)}$ is a tetrad $\vec{e}_{(a')}$ say, i.e. when we set
\begin{equation}
\vec{e}_{(a')}=\Lambda_{(a')}^{\phantom{a}(b)}\vec{e_{(b)}}
\end{equation}
then
\begin{equation}
\vec{e}_{(a')}\cdot\vec{e}_{(a')}=\eta_{(a)(a')}.	
\end{equation}

Let the CGO observer pass by the SO with momentarily velocity $V\equiv u^{(3)}/u^{(0)}$ where $u^{(i)}$ are components of CGO velocity vector in SO frame, i.e.
\begin{equation}
\vec{u}=u^{(i)}\vec{e}_{(i)}.
\end{equation}
Momentarily Lorentz transformation between CGO and SO now reads
\begin{equation}
\left[\Lambda_{(a')}^{(b)}\right]=\left[
\begin{array}{cccc}
\gamma & 0 & 0 & -\gamma\,V\\
0 & 1 & 0 & 0 \\
0 & 0 & 1 & 0 \\
-\gamma\, V & 0 & 0 & \gamma	
\end{array}
\right]
\end{equation}
where is
\begin{equation}
\gamma=(1-V^2)^{-1/2}
\end{equation}
with velocity parameter
\begin{equation}
V=\sqrt{\frac{h_2(r_{c})}{f(r_{c})}}\Omega_{c}
\end{equation}
where the angular frequency $\Omega_{c}$ is given by formula
\begin{eqnarray}
\Omega^2_{c}&=&\frac{u^\phi}{u^t}=\frac{f'(r_c)}{h'_2(r_c)}\nonumber\\
&=&\frac{M/r^2+B^2M/3+\mathcal{E}(r-2M)}{r\left[1-(B^2-2\mathcal{E})M^2-2(B^2-3\mathcal{E})r^2/3\right]}.
\end{eqnarray}
The tetrad of CGO relative to coordinate basis $\vec{e}_\mu$ reads
\begin{eqnarray}
\vec{e}_{(0')}&=&\frac{1}{\sqrt{1-V^2}}\left(\frac{1}{\sqrt{f}}\vec{e}_t-\frac{V}{\sqrt{h+2}}\vec{e}_\phi\right),\\
\vec{e}_{(1')}&=&\sqrt{f}\vec{e}_r,\\
\vec{e}_{(2')}&=&A\,\vec{e}_t + B\,\vec{e}_r+C\,\vec{e}_\theta,\\
\vec{e}_{(3')}&=&\frac{1}{\sqrt{1-V^2}}\left(-\frac{V}{\sqrt{f}}\vec{e}_t+\frac{1}{\sqrt{h_2}}\vec{e}_{\phi}\right).
\end{eqnarray}
\subsection{Initial Conditions}
Using tetrads that we have just determined, we can easily specify initial conditions for propagation vector components  $k^\mu$. They read
\begin{equation}
k^\mu=L_{(a)}^{\mu}k^{(a)}
\end{equation}
where the transformation matrix $L_{(a)}^{\mu}$ between SO and coordinate basis is written in form
\begin{equation}
\left[L_{(a)}^{\mu}\right]=\left[
\begin{array}{cccc}
\frac{1}{\sqrt{f}} & 0 & 0 & 0\\
0 & \sqrt{f} & 0 & 0\\
A & B & C & 0\\
0 & 0 & 0 & \frac{1}{\sqrt{h_2}}
\end{array}
\right]
\end{equation}
while the transformation matrix between CGO and coordinate basis reads
\begin{equation}
\left[L_{(a)}^{\mu}\right]=\left[
\begin{array}{cccc}
\gamma\frac{1}{\sqrt{f}} & 0 & 0 & -\gamma\,V\frac{1}{\sqrt{h_2}}\\
0 & \sqrt{f} & 0 & 0\\
A & B & C & 0\\
-\gamma\,V\frac{1}{\sqrt{f}} & 0 & 0 & \gamma\frac{1}{\sqrt{h_2}}
\end{array}
\right].
\end{equation}

Initial propagation vector components $k^\mu_0$ are finally determined by formulae:
\begin{itemize}
	\item \emph{SO frame}
	\begin{eqnarray}
	k_0^t&=&\frac{k^{(0)}}{\sqrt{f}}\left(1+\frac{w}{\sqrt{H}}\frac{k^{(2)}}{k^{(0)}}\right),\\
	k_0^r&=&k^{(0)}\sqrt{f}\left(\frac{k^{(1)}}{k^{(0)}}-\frac{w}{\sqrt{H}}\frac{k^{(2)}}{k^{(0)}}\right),\\
	k_0^\theta&=& k^{(0)}\frac{2\sqrt{f}}{\sqrt{H}}\frac{k^{(2)}}{k^{(0)}},\\
	k_0^\phi&=&\frac{k^{(0)}}{\sqrt{h_2}}\frac{k^{(3)}}{k^{(0)}}
	\end{eqnarray}
	\item \emph{CGO frame}
	\begin{eqnarray}
	k_0^t&=&\frac{k^{(0)}}{\sqrt{f}}\left(\gamma+\frac{w}{\sqrt{H}}\frac{k^{(2)}}{k^{(0)}}-\gamma V \frac{k^{(3)}}{k^{(0)}}\right),\\
	k_0^r&=&k^{(0)}\sqrt{f}\left(\frac{k^{(1)}}{k^{(0)}}-\frac{w}{\sqrt{H}}\frac{k^{(2)}}{k^{(0)}}\right),\\
	k_0^\theta&=& k^{(0)}\frac{2\sqrt{f}}{\sqrt{H}}\frac{k^{(2)}}{k^{(0)}},\\
	k_0^\phi&=&\frac{\gamma}{\sqrt{h_2}}k^{(0)}\left(-V+\frac{k^{(3)}}{k^{(0)}}\right)
	\end{eqnarray}
\end{itemize}
\subsection{Results}
\begin{figure}[h]
	\begin{center}
		\begin{tabular}{c}
			\includegraphics[scale=0.5]{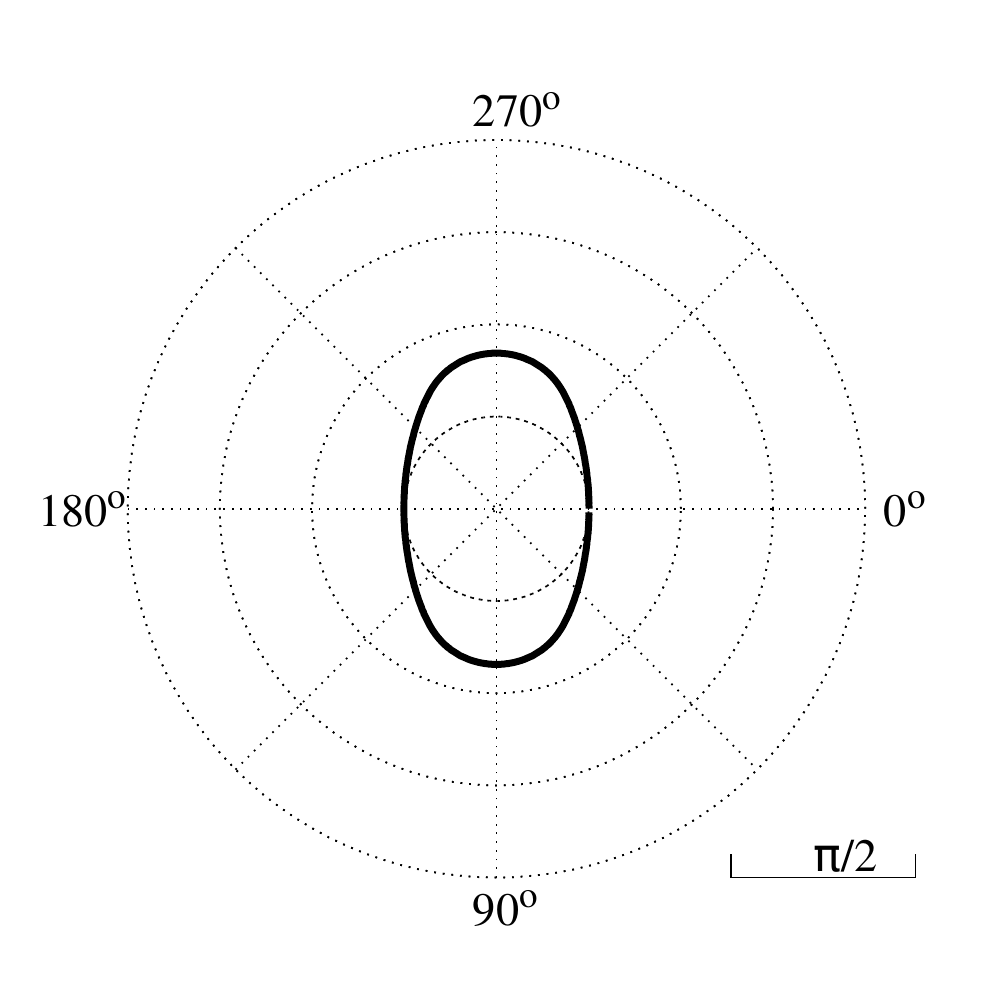}\\
			\includegraphics[scale=0.5]{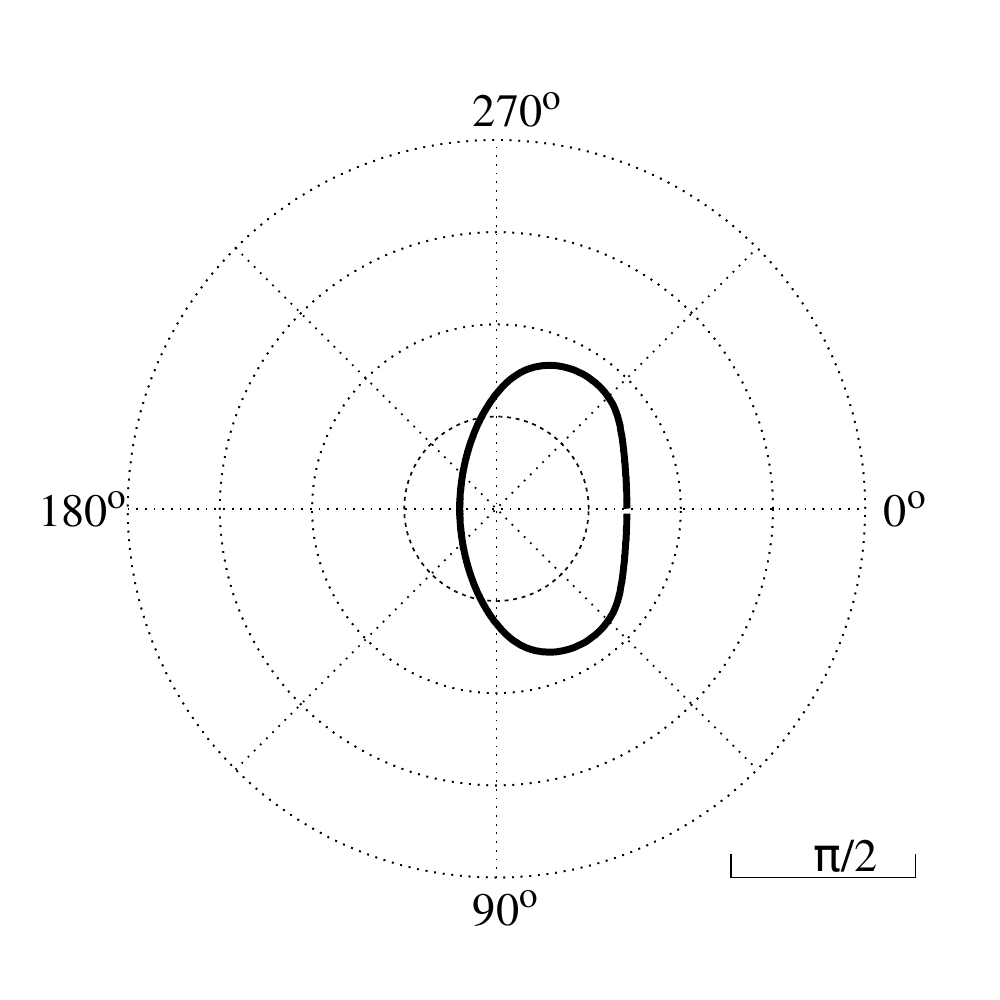}
		\end{tabular}
		\caption{{The shadow of P-P black hole seen by SF (top) and COF (bottom) observers orbiting central body in equatorial plane at radial coordinate $r_o=6M$. The black hole parameters are set to $(\mathcal{B},\mathcal{E})=(10^{-2},10^{-3})$ and $r_0=10M$.}\label{fig_apndx_1}}
	\end{center}
\end{figure}

\begin{figure}[h]
	\begin{center}
		\begin{tabular}{c}
			\includegraphics[scale=0.5]{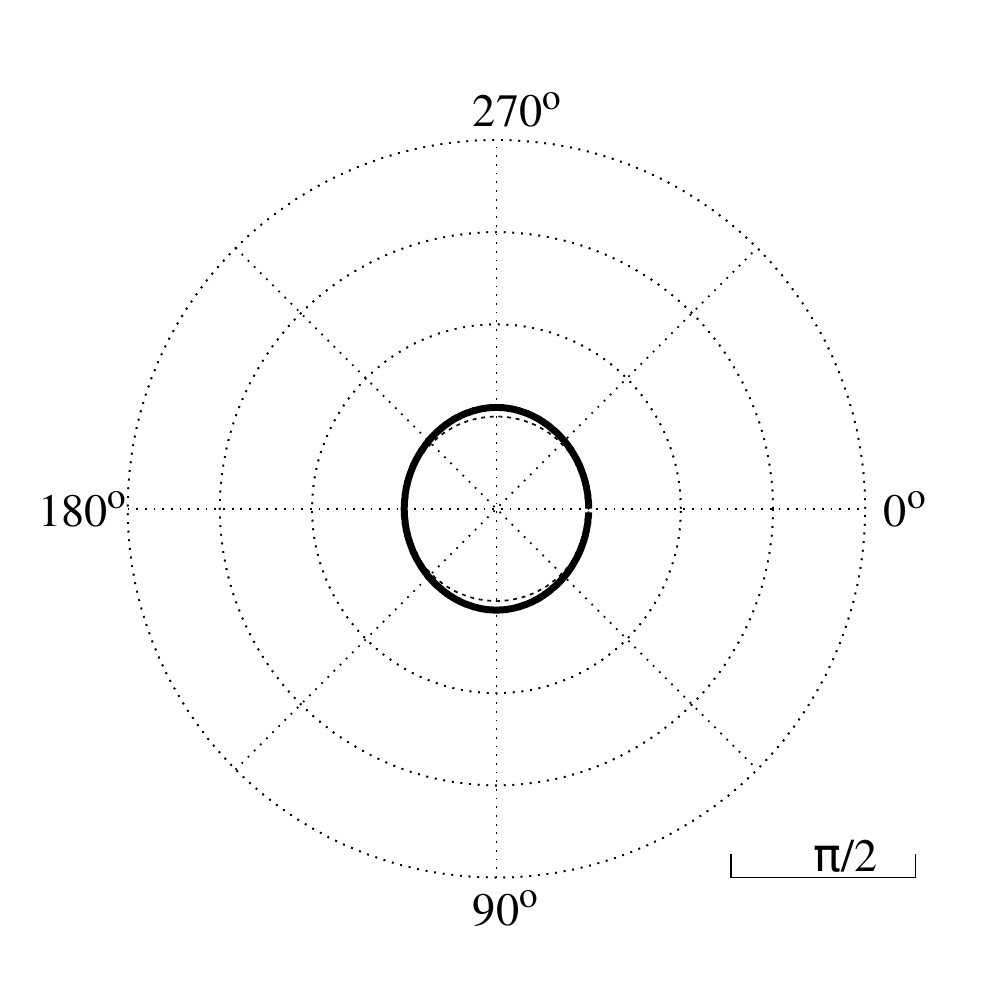}\\
			\includegraphics[scale=0.5]{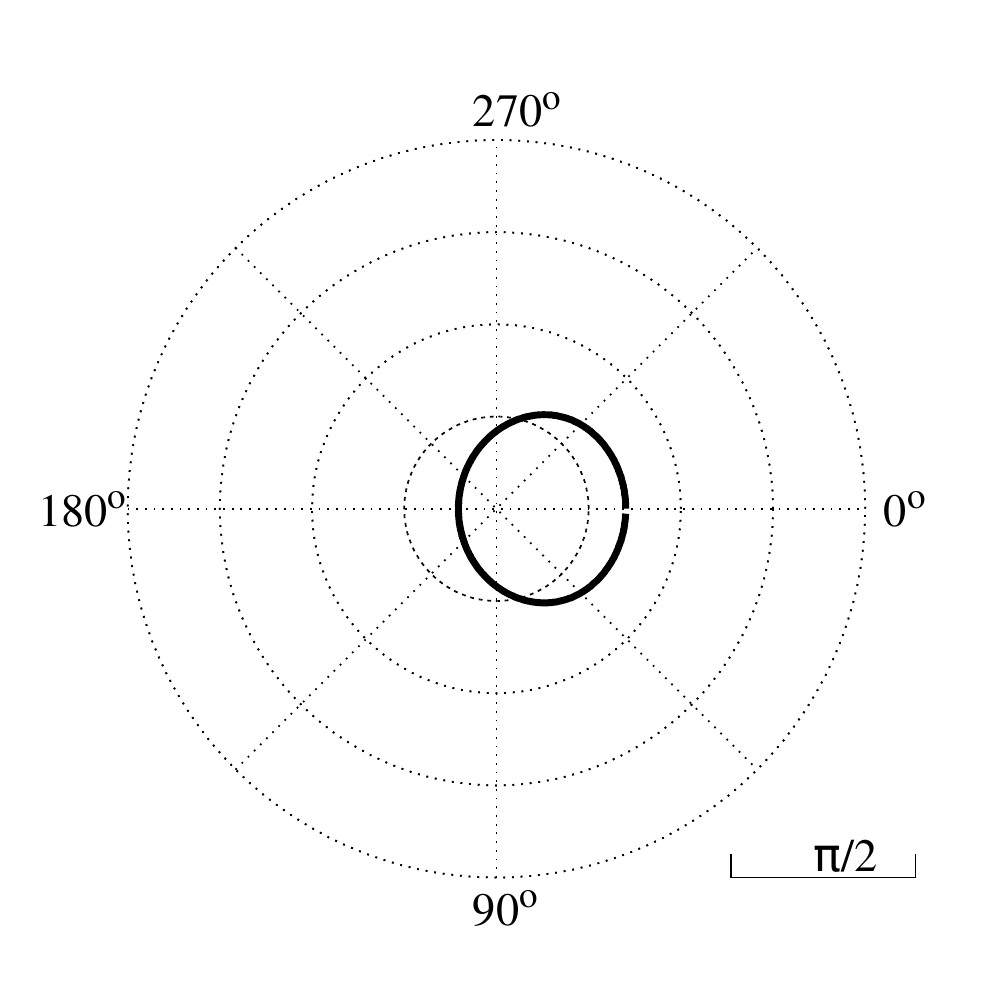}
		\end{tabular}
		\caption{{The shadow of P-P black hole seen by SF (top) and COF (bottom) observers orbiting central body in equatorial plane at radial coordinate $r_o=6M$. The black hole parameters are set to $(\mathcal{B},\mathcal{E})=(10^{-3},10^{-4})$ and $r_0=10M$.}\label{fig_apndx_2}}
	\end{center}
\end{figure}

Having initial conditions for photon propagation vector, one can integrate geodesic equation and determine the boundary of the black hole shadow, using procedure discussed in Section V. We constructed the shadows for two illustrative examples of P-P black hole having parameters $(\mathcal{B},\mathcal{E})=(10^{-2},10^{-3})$ (see Fig. \ref{fig_apndx_1}) and $(\mathcal{B},\mathcal{E})=(10^{-3},10^{-4})$ (see Fig. \ref{fig_apndx_2}).

\begin{figure}[H]
	\begin{center}
			\includegraphics[scale=0.5]{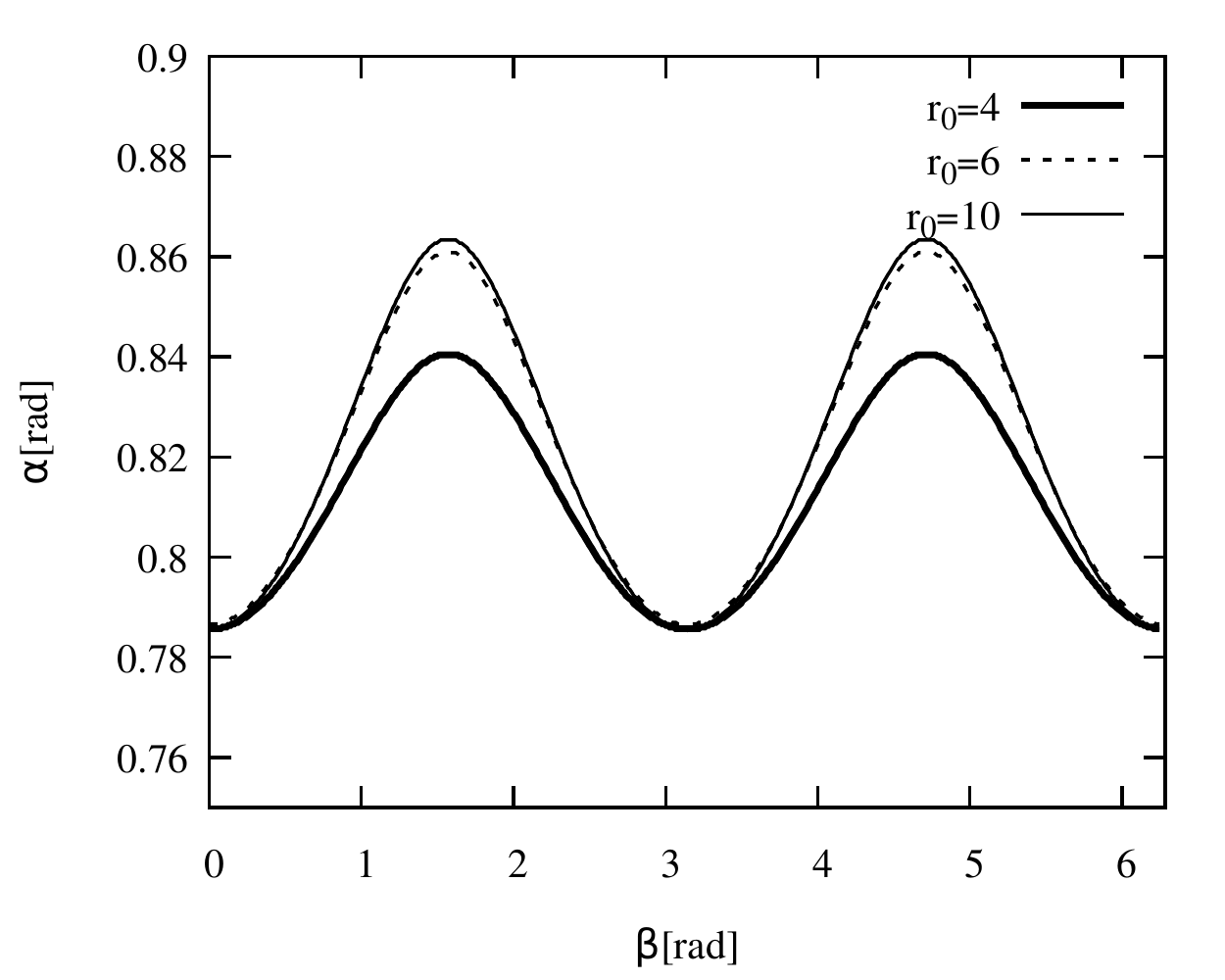}
		\caption{The shadow of P-P black hole observed by SF. The only parameter varied is $r_0=4$, $6$, $10$. The spacetime parameters are $(\mathcal{B},\mathcal{E})=(10^{-3},10^{-4})$.\label{fig_apndx_3a}}
	\end{center}
\end{figure}

\begin{figure}[H]
	\begin{center}
			\includegraphics[scale=0.5]{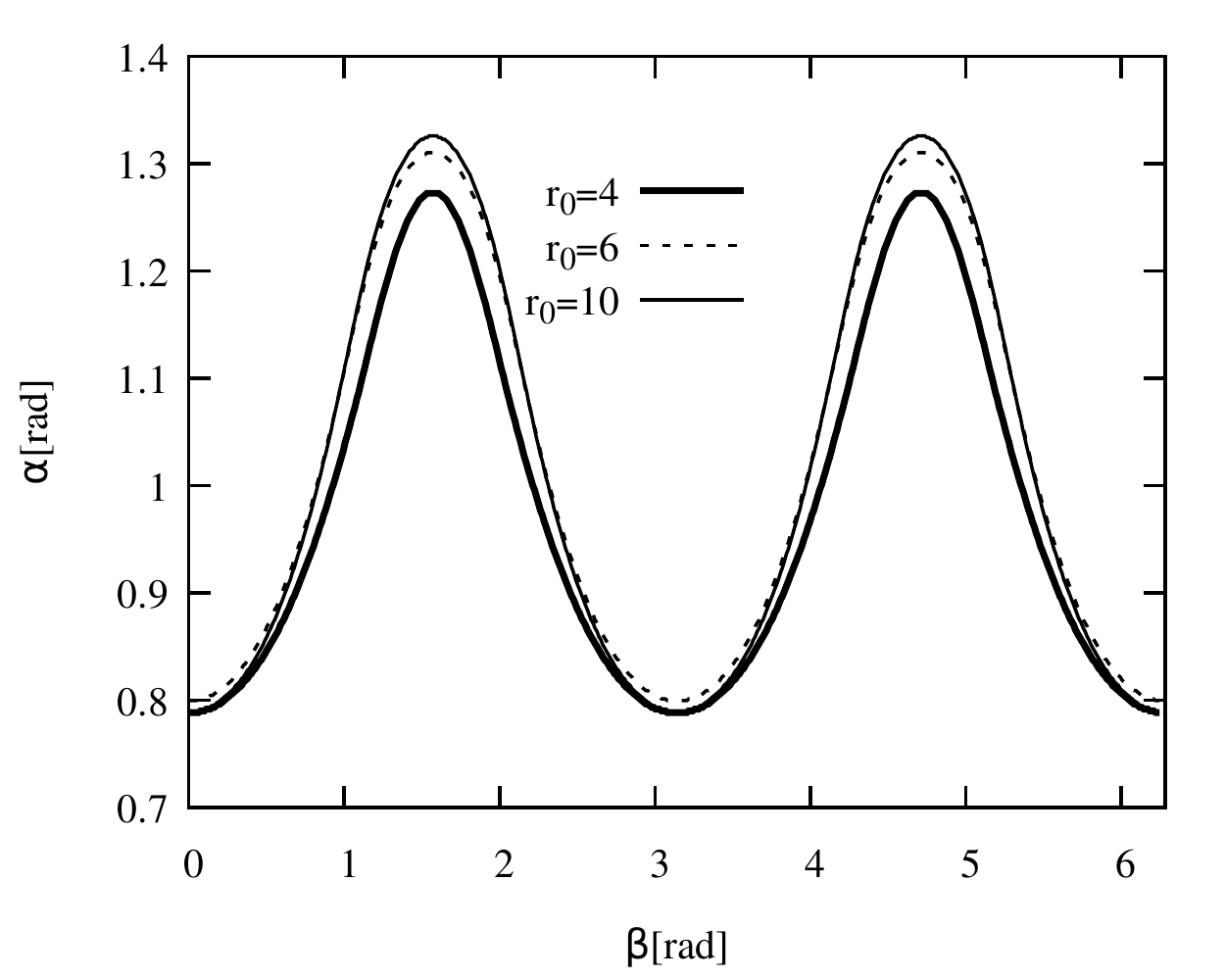}
		\caption{The shadow of P-P black hole observed by SF. The only parameter varied is $r_0=4$, $6$, $10$. The spacetime parameters are  $(\mathcal{B},\mathcal{E})=(10^{-2},10^{-3})$ .\label{fig_apndx_3b}}
	\end{center}
\end{figure}

The static observer in the vicinity of P-P black hole sees the same kind of black bole shadow deformation relative to Schwarzschild black hole shadow as it is measured by the distant observer. When one increases the strength of tidal effects from the torus and considers an observer on circular Keplerian orbit another, Kerr-like deformation of the shadow is generated (see Figs.\ref{fig_apndx_2}), but this effect is irrelevant for distant observer.
We also illustrate that the choice of the radius $r_0$, where we tailor P-P spacetime on Schwarzschild spacetime,  does not influence the shape of the shadow provided $r_0>\mathrm{Max(r_{ph}(\theta))}$ (see Figs. \ref{fig_apndx_3a} and \ref{fig_apndx_3b}). The location of the tailor radius $r_0$ modifies the size of the P-P black hole shadow but keeps its nature. The closer we move $r_0$ to $r_{ph}$ the smaller is the effect of deformation (it is expected behavior because we are in fact replacing P-P for Schwarzschild spacetime).

\end{document}